\documentclass[preprint,12pt]{aastex}
\begin{document}
\newcommand{\Halpha}{H$\alpha$ }
\newcommand{\etal}{\mbox{et al.}}
\newcommand{\NHoo}{N_{\rm H}}
\newcommand{\NHxo}{N_{\rm x}}
\newcommand{\NHIo}{N_{\rm HI}}
\newcommand{\NHtwoone}{N$_{\rm 21cm}$ }
\newcommand{\NHII}{N_{\rm HII}}
\newcommand{\NHgx}{N_{\rm G}}
\newcommand{\acmcm}{cm$^{-2}$}
\newcommand{\acmcmcm}{cm$^{-3}$}
\newcommand{\as}{s$^{-1}$}
\newcommand{\Htwo}{H$_{2}$}
\newcommand{\LHB}{{\sc lhb}}
\newcommand{\ISM}{{\sc ism}}
\newcommand{\LISM}{{\sc lism}}
\newcommand{\MER}{{\sc mer}}
\newcommand{\IRAS}{{\it IRAS}}
\newcommand{\ROSAT}{{\it ROSAT}}
\newcommand{\COBE}{{\it COBE}}
\newcommand{\DIRBE}{{\it DIRBE}}
\newcommand{\EUVE}{{\it EUVE}}
\newcommand{\pp}{\phn}
\newcommand{\ppp}{\phn\phn}
\newcommand{\pppp}{\phn\phn\phn}
\newcommand{\pq}{\,$\pm$\,}
\newcommand{\pd}{\phn\phn\phn\,---}
\newcommand{\plong}{\hspace{10pt}}
\newcommand{\gte}{$\infty$\phn}
\newcommand{\ZY}{0.3,1.0}
\newcommand{\ZZ}{0.5,1.0}
\newcommand{\Msun}{$M_{\odot}$}
\newcommand{\ayr}{y$^{-1}$}
\newcommand{\pnt}{\phantom{-}}
\newcommand{\lan}{\langle}
\newcommand{\ran}{\rangle}
\newcommand{\ARAA}[2]{ARA\&A, #1, #2}
\newcommand{\ApJ}[2]{ApJ, #1, #2}
\newcommand{\ApJL}[2]{ApJL, #1, #2}
\newcommand{\ApJSS}[2]{ApJS, #1, #2}
\newcommand{\AandA}[2]{A\&A, #1, #2}
\newcommand{\AandASS}[2]{A\&AS, #1, #2}
\newcommand{\AJ}[2]{AJ, #1, #2}
\newcommand{\BAAS}[2]{BAAS, #1, #2}
\newcommand{\ASP}[2]{ASP Conf.\ Ser., #1, #2}
\newcommand{\JCP}[2]{J.\ Comp.\ Phys., #1, #2}
\newcommand{\MNRAS}[2]{MNRAS, #1, #2}
\newcommand{\N}[2]{Nature, #1, #2}
\newcommand{\PASJ}[2]{PASJ, #1, #2}
\newcommand{\RPP}[2]{Rep.\ Prog.\ Phys., #1, #2}
\newcommand{\ZA}[2]{Z.\ Astrophs., #1, #2}
\newcommand{\tenup}[1]{\times 10^{#1}}
\newcommand{\tu}[1]{\times 10^{#1}}

\title{Maximum Entropy Reconstruction of the Interstellar Medium:\\ I. Theory}
\author{John S.\ Arabadjis and Joel N.\ Bregman}
\affil{University of Michigan}
\affil{Ann Arbor, MI 48109-1090}
\affil{jsa@astro.lsa.umich.edu}
\affil{jbregman@astro.lsa.umich.edu}

\begin{abstract} 

We have developed a technique to map the three-dimensional structure of the
local interstellar medium using a maximum entropy reconstruction technique.  A
set of column densities ${\bf N}$ to stars of known distance can in principle
be used to recover a three-dimensional density field ${\bf n}$, since the two
quantities are related by simple geometry through the equation
${\bf N} = {\sf C} \cdot {\bf n}$, where ${\sf C}$ is a matrix characterizing
the stellar spatial distribution.  In practice, however, there is an infinte
number of solutions to this equation.  We use a maximum entropy reconstruction
algorithm to find the density field containing the least information which is
consistent with the observations.  The solution obtained with this technique
is, in some sense, the model containing the minimum structure.  We apply
the algorithm to several simulated data sets to demonstrate its feasibility and
success at recovering ``real'' density contrasts.

This technique can be applied to any set of column densities whose end points
are specified.  In a subsequent paper we shall describe the application of
this method to a set of stellar color excesses to derive a map of the dust
distribution, and to soft X-ray absorption columns to hot stars to derive a map
of the {\it total} density of the interstellar medium.

\end{abstract}


\section{Introduction} 

Given a set of projected density measurements of columns with known end points,
one can, at least in principle, recover the three dimensional distribution of
absorbing material.  One way to accomplish this deprojection is to simply draw
isocolumn contours \citep{sfeir,welsh}.  Here we propose an alternative
technique -- maximum entropy reconstruction (\MER) -- and apply a computer
implementation of the algorithm to simulated data sets.  We test its ability
to recover structure in the interstellar medium (\ISM) by varying several
parameters and the functional form of the entropy.  We then briefly sketch its
application to a number of archived data sets, to be presented in a subsequent
paper.

\section{Maximum Entropy Technique} 

The column or projected density of absorbing material $N({\bf x})$ along the
line of sight between an observer and a star at position ${\bf r}_{\star}$ is
defined as

\begin{equation}
N({\bf r}_{\star}) = \int_0^{r_{\star}} n({\bf x}) \, \, dr
\label{eq01}
\end{equation}

\noindent where $n({\bf x})$ is the number density of absorbers at position
${\bf x}$, usually measured in \acmcmcm.  It is $n({\bf x})$, the map of the
\ISM, that we seek.  Although the inversion of equation~\ref{eq01} is formally
trivial, we know only $N/r_{\star}$, and so there is an infinite number of
functions $n({\bf x})$ which satisfy the observed $N({\bf x}_{\star})$.  We
therefore take a discrete approach, approximating the \ISM\ as a coarse-binned
density field.

Consider a set of data consisting of $N_{\star}$ stars of known distance, to
which accurate intervening column densities have been measured.  We write the
set of column densities to these $N_{\star}$ stars as the vector
${\bf N} = (N_1, N_2, \ldots N_{N_{\star}})$.  Let us divide the space occupied
by these stars into $N_c$ cells of size $s$.  (In practice the cells need not
be cubic -- more generally we have $s_x$, $s_y$ and $s_z$.  For example, it is
sometimes useful to set $s_z$ to a large number to build a planar model of the
\ISM.)  The model we seek consists of the set of number densities of these
cells (${\bf n}$, a vector of length $N_c$) which will reproduce the observed
column densities.  The observations form a constraint equation

\begin{equation}
{\bf N} = {\sf C} \cdot {\bf n} \, ,
\label{eq02}
\end{equation}

\noindent where ${\sf C}$ is a matrix with $N_c$ rows and $N_{\star}$ columns
describing the line-of-sight/cell intersections for each star.

For the moment, let us assume that the observer is located at the corner shared
by 8 cells.  (In the actual implementation of this algorithm we place the
observer at the center of a cell, however, to enhance the numerical stability
of several matrix inversions.)  If $s$ is small enough, no two lines of sight
will intersect the same cell, and there will be an infinite number of solutions
to this equation.  It is clear, however, that some solutions will be preferred
over others.  For example, let us imagine that our data set consists of only
one stellar position and intervening column density.  It is easy to imagine two
different models -- the minimum and maximum structure configurations
(Figure~\ref{f01}) -- which reproduce the observations.  Figure~\ref{f01}a
shows the material spread out evenly along the line of sight, whereas in
Figure~\ref{f01}b the material inhabits only one cell.  In the absence of any
additional information both models reproduce the observations equally well.
Information theory tells us that the former is to be preferred to the latter,
however, since it makes the fewest assumptions about the underlying structure
-- i.e.\ it contains less information.  Since our goal is to uncover structure
in the \ISM, the most conservative approach is to find the solution containing
the minimum structure.  Thus we seek a configuration which is a solution to
equation~\ref{eq02} and which contains the minimum amount of information.

The quantity of information contained in a map is usually measured using the
Shannon entropy \citep{shannon,kapur}.  To construct the map with the fewest
assumptions about the underlying structure, but consistent with a set of
observations, we invoke Jaynes' principle of maximum entropy \citep{jaynes}.
The nomenclature ``entropy'' was adopted because the Shannon entropy was shown
by \citet{jaynes} to be identical to the classical Boltzmann entropy.

In general, a map can be characterized by a distribution function
$f = f(t,{\bf x},{\bf v})$, where $f$ is the mass per unit volume per velocity
interval.  The amount of information contained in the distribution is
characterized by the  Boltzmann $H$ function,

\begin{equation}
H = \int f \log{f} \ d{\bf x} \ d{\bf v}
\label{eq03}
\end{equation}

\noindent which is related to the thermodynamic entropy $S$ through

\begin{equation}
H = -\frac{S}{kV}
\label{eq04}
\end{equation}

\noindent Here V is the volume covered by the map, and $k$ is Boltzmann's
constant.

The choice of grid resolution is fundamental to the character of the
reconstructed map.  We will choose a cell size $s$ such that
$v_{\rm ISM} \cdot t_{\rm obs} \ll s$, where $v_{\rm ISM}$ is the typical
velocity of material in the \ISM\ and $t_{\rm obs}$ is the timescale
\mbox{spanned} by the observations.  Thus we can ignore any time dependence in
the phase space density, making it a function of \mbox{cell} position only.  In
addition, $v_{\rm ISM} \ll c$ throughout the Galactic \ISM, and so the
absorptive properties of the material in each cell are independent of the
velocity of the material.  Thus we replace $f({\bf x},{\bf v}, t)$ with the
coarse-binned $\rho_i$.  Scaling the entropy by the mean particle mass and
Boltzmann's constant we obtain

\begin{equation}
S = -\sum^{N_c}_{i=1} n_i \log{n_i}
\label{eq05}
\end{equation}

\noindent where $n_i$ is the number density of absorbers in cell $i$.  The
analogy between $n_i$ and probability $p_i$ of finding a particle in cell $i$
leads immediately to the identification of equation~\ref{eq05} as the Shannon
entropy.  We shall, however, refer to this form as the ``Boltzmann entropy'' of
the map, in keeping with past astrophysical applications of the maximum entropy
principle.

In the present problem we are not restricted to using the Boltzmann entropy.
Boltzmann's $H$ theorem was originally formulated to show that the
thermodynamic entropy is the only non-decreasing function of the time-dependent
distribution function for an ideal gas \citep{boltzmann,tolman}.  In
collisionless systems, however, there is an infinite number of $H$ functions
which exhibits this behavior \citep{tremaine}.  These are of the form

\begin{equation}
S = -\sum^{N_c}_{i=1} C_i(n_i)
\label{eq06}
\end{equation}

\noindent where $C$ is any convex function of the distribution function.
Following \citet{richstone} we shall use the term ``entropy'' to refer to any
function of this form, although, in the context of information, ``uncertainty''
might be a more appropriate term, as \citet{kapur} point out.

The \ISM\ is an extremely complex physical system.  Although it can be
considered a ``collisional'' system in some sense, astrophysical processes such
as radiative cooling, energy injection by supernovae, and self-regulating star
formation render this characterization of its self-interaction too simplistic.
Thus there are no compelling arguments based upon first principles which favor
one form of the entropy over another.  We shall instead turn to numerical
experimentation and convenience to guide our choice of the entropy.  We shall
return to this topic in Section~\ref{sec_entropy}.

Regardless of the form of entropy we ultimately adopt, we seek seek the
solution of equation~\ref{eq02} where $S$ is a maximum, or, equivalently,
we seek the map containing the minimum information.  (Although there are
many configurations wherein $S$ is minimized, there is only one, for a given
grid, where it is maximized.) This problem is readily solved using the method
of Lagrange multipliers \citep{jaynes}.  For example, \citet{richstone} adapted
this technique to solve the problem in a stellar dynamical context.  We adopt
this approach here, recasting it for the problem at hand.

The (generally nonlinear) variational equation we must solve is

\begin{equation}
\delta S - \delta{\bf N}\cdot {\bf L} = 0
\label{eq07}
\end{equation}

\noindent where the vector ${\bf L} = (L_1,L_2,\ldots,L_{N_c})$ is the set of
Lagrange multipliers to be determined later.  In the limit of small variations
we have

\begin{equation}
\nabla S - (\nabla{\bf N})^t \cdot {\bf L} = 0
\label{eq08}
\end{equation}

\noindent where the operator $\nabla$ is defined as

\begin{equation}
\nabla = {\bf \hat{e}}_1 \frac{\partial}{\partial n_1} +
{\bf \hat{e}}_2 \frac{\partial}{\partial n_2} + \ldots +
{\bf \hat{e}}_{N_c} \frac{\partial}{\partial n_{N_c}}
\label{eq09}
\end{equation}

\noindent and $(\nabla{\bf N})^t$ is the transpose of $\nabla{\bf N}$.  Since
the column density is related to the number density through
equation~\ref{eq02}, $\nabla {\bf N} = {\sf C}$, and we have then

\begin{equation}
\nabla S - {\sf C}^t \cdot {\bf L} = 0
\label{eq10}
\end{equation}

To solve this equation we guess a solution ${\bf n}_0$ and iterate using
the Newton-Raphson method.  Expanding about the initial guess and keeping
terms only to first order yields

\begin{equation}
\nabla S|_{\rm new} = \nabla S|_{{\bf n}_0} + \nabla (\nabla S|_{{\bf n}_0})
\cdot \delta {\bf n}
\label{eq11}
\end{equation}

\noindent Inserting equation~\ref{eq11} into equation~\ref{eq10}, and solving
for $\delta {\bf n}$ yields

\begin{equation}
\delta {\bf n} = (\diamondsuit S|_{{\bf n}_0})^{-1} \cdot {\sf C}^t \cdot
{\bf L} - (\diamondsuit S|_{{\bf n}_0})^{-1} \cdot \nabla S|_{{\bf n}_0}
\label{eq12}
\end{equation}

\noindent where $\diamondsuit S \equiv \nabla(\nabla S)$.  To make use of this
equation we must first calculate the Lagrange multipliers.  Dotting both sides
with ${\sf C}$ and solving for ${\bf L}$ we obtain

\begin{equation}
{\bf L} = ({\sf C} \cdot \diamondsuit S|_{{\bf n}_0}^{-1} \cdot {\sf C}^t)^{-1}
\cdot \delta {\bf N} + \, 
( {\sf C} \cdot \diamondsuit S|_{{\bf n}_0}^{-1} \cdot {\sf C}^{t} )^{-1}
\cdot {\sf C} \cdot \diamondsuit S|_{{\bf n}_0}^{-1} \cdot
\nabla S|_{{\bf n}_0}
\label{eq13}
\end{equation}

\noindent Here we have identified ${\sf C} \cdot \delta{\bf n}_0$ as
$\delta {\bf N}$, the difference between the observations and the model column
densities corresponding to the guessed solution ${\bf n}_0$.
Equation~\ref{eq12} is applied iteratively, with the Lagrange multipliers
calculated anew at each step using equation~\ref{eq13}, until a satisfactory
solution is obtained.  The computational expense is dominated by the inversion
of $\diamondsuit S$, an $N_c\times N_c$ matrix.

The density of material in each cell is physically (though not mathematically)
restricted to non-negative values.  In order to impose this condition upon our
solution we require that each component of $\delta {\bf n}$, as calculated in
the code by equation~\ref{eq12}, be no greater in magnitude than the current
value of density in that cell.  This guarantees that each cell has a
non-negative density at every step in the iterative reconstruction and in the
final maximum entropy solution.

\section{Simulated Data Sets} 

When we apply the \MER\ algorithm to a real data set, we will not know whether
the resultant map is close to the actual \ISM\ structure.  (We can be sure that
the reconstructed map is very different from the true structure if the
iteration scheme fails to converge, however.)  To ascertain how well the
algorithm reconstructs underlying distribution of absorbing matter we must test
its performance upon simulated data sets.  We postulate a 3D density structure
$n({\bf x})$ for the \ISM\ and a stellar spatial distribution, and calculate
$N({\bf x}_{\star})$ for each star using equation~\ref{eq01}.  We then apply
the \MER\ algorithm to the faked data set ${\bf N}({\bf x}_{\star})$ and
compare the maximum entropy configuration with the simulated density structure.

There are three aspects of the scheme which need to be explored using these
simulated data.  These are (1) the form of the entropy used in the maximization
routine, (2) the size of the spatial cells used, for a given sampling density,
and (3) the sampling density of the stars in the data set, compared
with the scale length of the smallest structures in the \ISM.

\subsection{Functional Form of the Entropy\label{sec_entropy}}

A strict interpretation of information theory dictates that the entropy we
should maximize is the Shannon entropy (i.e.\ the classical Boltzmann entropy),
in the form given by equation~\ref{eq05}.  However, in the spirit of
experimentation, we also ran tests for three other forms of the entropy as
well.  Two of these are ``true'' entropies in the the sense of
\citet{tremaine}, following the form of equation~\ref{eq06}.  They are

\begin{equation}
S = -\sum^{N_c}_{i=1} \, n_i^2
\label{eq14}
\end{equation}
\noindent and
\begin{equation}
S = -\sum^{N_c}_{i=1} \, e^{n_i}
\label{eq15}
\end{equation}

\noindent The former, because it is a simple quadratic, should converge in
only one iteration in the Newton-Raphson scheme.  The fourth function we tried
was

\begin{equation}
S = \sum^{N_c}_{i=1} \, n_i \, e^{-n_i}
\label{eq16}
\end{equation}

\noindent This function is not of the form given by equation~\ref{eq06}, and so
we refer to it as a ``pseudo-entropy''.  Although it is not of the form
discussed by \citet{tolman} and \citet{tremaine}, like these other forms it
possesses a single global maximum -- a property which, as it turns out, is
probably sufficient to guarantee its utility.  The four entropy types are
compared in Figure~\ref{f02}.

The simulated density field used for this experiment is shown in
Figure~\ref{f03}.  It consists of an ambient medium with a density of
$\sim 0.1$ \acmcmcm, with two structures superimposed upon it.  One is a wall
of material with a gaussian amplitude $\sim 150$ \acmcmcm; the other is a small
cloud with gaussian amplitude $\sim 70$ \acmcmcm.  The entire map has
dimensions of 1017.5 pc $\times$ 632.5 pc.

Throughout this density field we distribute 458 stars in a random fashion (but
with a minimum separation) and use the calculated columns to try to reconstruct
the field.  \MER\ maps for the four entropy types given by
equations~\ref{eq05}, \ref{eq14}, \ref{eq15}, and \ref{eq16} are shown in
Figure~\ref{f04}.  Each reconstruction uses a cell size of 27.5 pc $\times$
27.5 pc, for a total grid size of $37 \times 23$.  Only 806 of the 851 cells
are active in the reconstruction, however, because many edge cells are not
pierced by a stellar line of sight.  (These cells, shown in dark gray, are
obvious toward the edges because they contain no stars.)  We have set a column
match tolerance of $10^{-4}$ for convergence termination, although in practice
this will be set by the column measurement accuracy.  (It should be noted that
the cell size was optimized for this example.)  Table~\ref{t01} lists the mean
column error, rms column error, the mean density error, and the rms density
error, for each of the reconstructions.  Quantities with the $mer$ subscript
obviously refer to properties of the reconstruction.  Column densities with no
subscript represent the input data, while number densities with no subscript
represent the density of the ``actual'' configuration at the cell center.

Two conclusions can be drawn from this experiment.  The first is that \MER\ is
an effective way to recover real density contrasts without introducing
significant spurious structure.  Both structures are clearly reproduced by
the algorithm with the appropriate amplitudes.  And in no case do we find
extraneous structure on scales larger than a pixel.

The second observation is that the \MER\ algorithm is quite insensitive to the
form of the entropy used.  The reconstructions created using the four different
entropy types are nearly identical; in fact, one would be hard-pressed to
distinguish one from another without referring to Table~\ref{t01}.  Even the
pixel-sized variations occur in the same location, suggesting that the
``noise'' in each reconstructed map is due to the peculiarities of the stellar
distribution with respect to the cell grid, rather than tied to a specific form
of the entropy.  We thefore conclude that it is best to use the quadratic
entropy, to save on computational expense.  If we normalize the time to
calculate the entropy and its derivatives to the that of the quadratic form,
we have $t_{\rm Boltz}=1.3$, $t_{\rm expo}=1.5$, and $t_{\rm pseudo}=2.0$.

In practice, program run-times are actually {\it much} shorter using the
quadratic entropy because the \MER\ algorithm usually converges in one or two
iterations, compared with four to seven for the other entropy types.  These
run-time savings are greatest when the grid contains a large number of cells
(and by extention a large number of stars, since they scale linearly in
optimized configurations).  In the 458-star simulation, a single iteration of
the code running on a 450 MHz Pentium II processor took 52 s when using a grid
of 684 cells, and 1230 s with a 2285-cell grid (see Section~\ref{sec_cell}).
So while using the quadratic entropy may only save a few minutes for a small
number of stars, the difference can be several hours for a large data set.

\subsection{Cell size\label{sec_cell}}

We next sought to understand the importance of cell size for a given stellar
sampling density.  The number density of the \ISM\ should really be calculated
differentially from the column data through equation~\ref{eq01}; the \MER\
algorithm, however, uses a differencing scheme.  Since the number density of
absorbers in a particular cell is constant across the cell in the reconstructed
map, we expect there to be numerical instabilities when the cell size is
comparable to or larger than the size scale of structures in the underlying
\ISM.  If the cell is so large that it contains two stars, and a non-negligible
density gradient, there may actually be no solution to the constraint equation,
regardless of any entropy considerations.

We examined the effect of cell size modulation on the reconstructed map, for a
fairly dense simulated stellar distribution.  For this set of experiments we
use the simulated density map shown in Figure~\ref{f05}.  It contains three
clouds of various size scale and maximum density.  From largest to smallest,
the gaussian amplitudes are $\sim 100$, 40, and 35 \acmcmcm.  The entire map
has dimensions of 1012.5 pc $\times$ 607.5 pc, with 666 stars distributed
throughout.  We varied the cell size from 27.5 pc on a side, down to 12.5 pc,
in increments of $-2.5$ pc.  These \MER\ maps are shown in Figure~\ref{f06}.

Table~\ref{t02} gives the error statistics for each reconstruction.  We omit
the 27.5 pc cell \MER\ map because the reconstruction failed -- the algorithm
was terminated after one iteration because the solution was divergent.  The
reason for this is that several cells contain two stars and a sufficiently
steep density gradient, resulting in zero solutions of the constraint equation.
Each of the reconstructions shown in Figure~\ref{f06} fit the data reasonably
well (the column tolerance was set to $10^{-3}$).  As the cell size decreases,
however, the solution begins to show radial striations as more and more cells
are intersected by only one or two stellar lines of sight.  In the limiting
case of infinitesimal cells, every active cell in the grid is pierced only
once; in essence, each individual line of sight reduces to the configuration
illustrated in Figure~\ref{f01}.  Thus material is smeared evenly along each
absorption path, and the $N_{\star}\times N_c$ optimization problem reduces to
$N_{\star}$ separate $\alpha_k N_c$ problems
($\sum^{N_\star}_{k=1} \alpha_k=1$).  Since no cell is intersected by more than
one line of sight, these paths no longer ``communicate'' with each other, and
their individual entropy contributions decouple during the optimization.

For this particular case the solution is optimized (in the sense of matching
the ``true'' 3D density) for a cell/star ratio of about 7/4.  However, barring
peculiarities in stellar position within each cell,
$\lan N_{mer}-N \ran \rightarrow 0$ as the cell size is reduced.  It would
be identically zero once each cell is pierced by only one stellar line of
sight, although the number density errors would grow as structure is washed out
along the line of sight.

\subsection{Sampling density\label{sec_sampling}}

Finally we examined the ability of the algorithm to detect \ISM\ structures
for a limited stellar sampling density.  Again we use the simulated \ISM\ of
Figure~\ref{f05}.  We ran twelve \MER\ reconstructions for $N_{\star}$ from
13 to 342, with optimized cell numbers from 22 to 342 (Figure~\ref{f07};
Table~\ref{t03}).  Using a column tolerance of $10^{-4}$, the cell size was
varied in increments of 5 pc to find the best reconstruction.  Again, dark gray
cells toward the edges which contain no stars are disregarded in the \MER.

The reconstruction using 13 stars can detect only the largest of the three
clouds.  The map is, however, quite good, considering that it gets the
amplitude right, and matches the columns to within 0.003\%.  The reconstruction
using 34 stars probably resolves the small cloud closest to the sun.
The 48-star map may actually resolve both small clouds, although it is not a
robust detection.  Although the column fit is nearly perfect, the map shows an
unfortunate feature common to low-density sampled maps -- the \MER\ algorithm
awards the pixel containing the sun an excess of material, at the expense of
the 8 pixels which surround it.  Thus if the goal of the reconstruction is to
study the {\it very} local \ISM, one cannot trust a high density contrast in
only the central (solar) bin.  By the time we get to 120 and 177 stars, both
small clouds are clearly detected, as is the low-density bridge between the two
largest clouds.  Clearly, the sample of stars must be sufficiently dense such
that the \ISM\ structures to be detected are at least as large as the mean
spacing between stars.

\section{Summary and applications} 

Maximum entropy reconstruction is a powerful technique for recovering the
structure of the \ISM\ from column density data sets.  By modulating the cell
size, one can find a solution which recovers the underlying structure in the
\ISM.  The lower limit to the size scale of detected structures is set by the
stellar sampling density.  The reconstruction algorithm is insensitive to the
particular form of the entropy employed, so it is most prudent to use
computationally simple and rapidly converging forms such as the quadratic
entropy (equation \ref{eq14}).

The next step is to apply this technique to existing astronomical data sets.
Different components of the \ISM\ can be traced by absorption and emission
measurments made at different wavelengths.  Using a variety of absorptive and
emissive tracers, we should be able to map regions of varying ionization and
molecular state and elemental composition, and compare their spatial
distributions.

We are currently working to map the distribution of dust in the \ISM\ using
column densities derived from stellar color excess measurements.  In addition,
we are using archived \ROSAT\ observations of hot stars to determine X-ray
absorption columns.  Since the soft X-ray cross section is insensitive to the
molecular and (modest) ionization state of the absorber \citep{arabadjis},
reconstructing the density field using X-ray columns will map the {\it total}
density of material the local \ISM.  Additionally, by exploiting the spectral
resolution of instruments aboard XMM-Newton and Chandra, one may be able to
separate this material by ionization state.  For example, $\Delta E/E \sim 50$
for the non-dispersive ACIS detector aboard Chandra, resulting in a spectral
resolution of about 10 eV at the prominent absorption edge of oxygen at 0.5
keV.  Since the neutral and first three ionization states of oxygen are
separated by about 20 eV each (the potentials are 13.6, 35.1, and 54.9 eV), it
may be possible to distill the oxygen distribution into its dominant ionization
states.

Several \ISM\ tracers also contain velocity information.  Optical absorption
lines in stellar spectra, originating from the \ISM, provide a measure of the
distribution of metals not bound up in dust grains, and maps constructed from
their columns may provide insight into the gas phase chemistry of \ISM.
Additionally, the velocity structure of a resolved absorption line can be used
in conjuction with a rotation curve and a second application of the maximum
entropy principle to further constrain the distribution of material along the
line of sight.  Similarly, velocity-resolved 21 cm emission maps can be used
to construct a distribution of neutral hydrogen along each line of sight.

The applicability of this technique to absorption measurements at almost any
wavelength comes at a particularly fortuitous time.  Most astronomical
satellite programs support an on-line archive containing their public domain
observations, many of which can be mined for projected density measurements.
With the number number of of these archives growing on an almost daily basis,
this technique should find wide application in the near future.

The authors would like to acknowledge support from NASA grant NAG5-3247.  JSA
would like to thank Doug Richstone for useful discussions.



\clearpage
\plotone{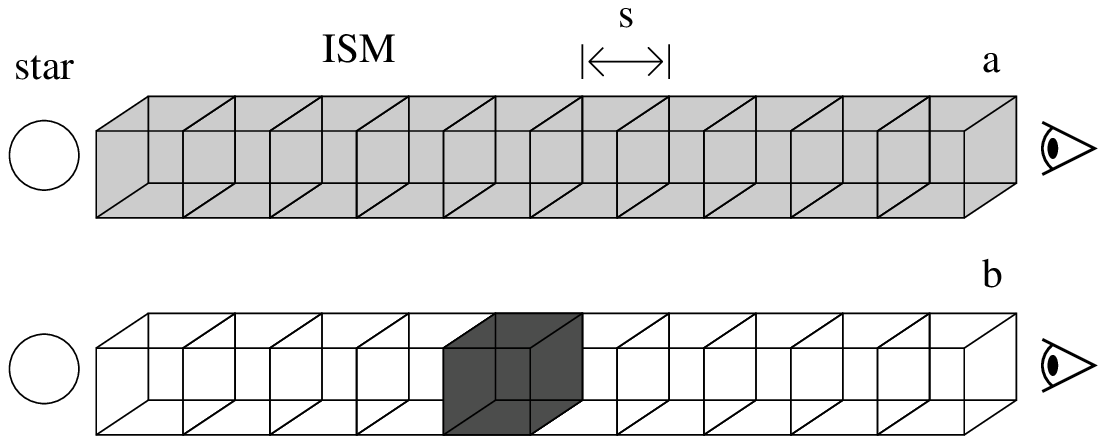}

\figcaption{Two discrete-bin \ISM\ models which result in the same intervening
column density toward a star.
\label{f01}}

\clearpage
\plotone{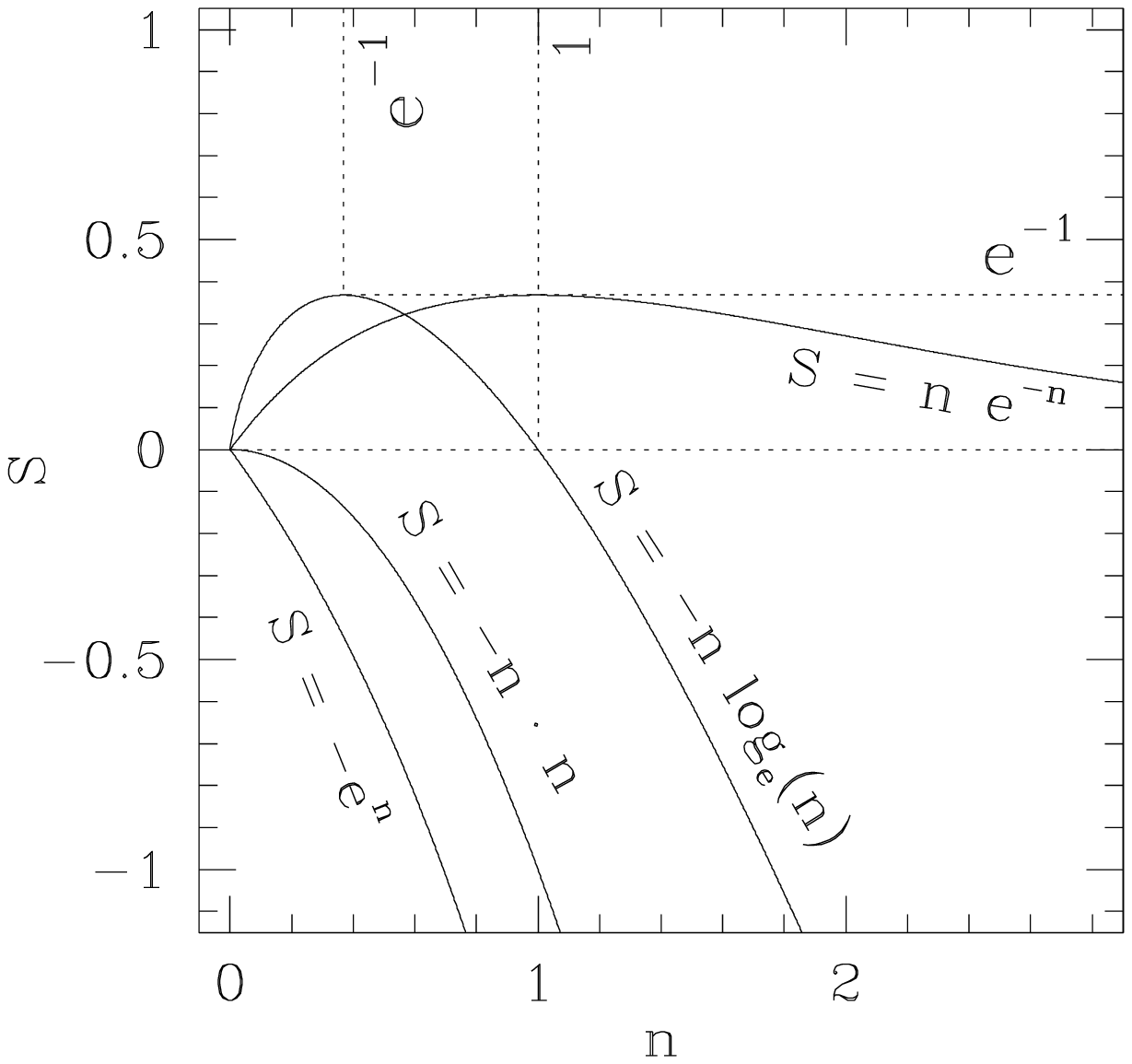}

\figcaption{The four entropy forms examined in this study.  Note the concavity
inflection of the pseudo-entropy $S=n e^{-n}$ at $n=2$.
\label{f02}}

\clearpage
\plotone{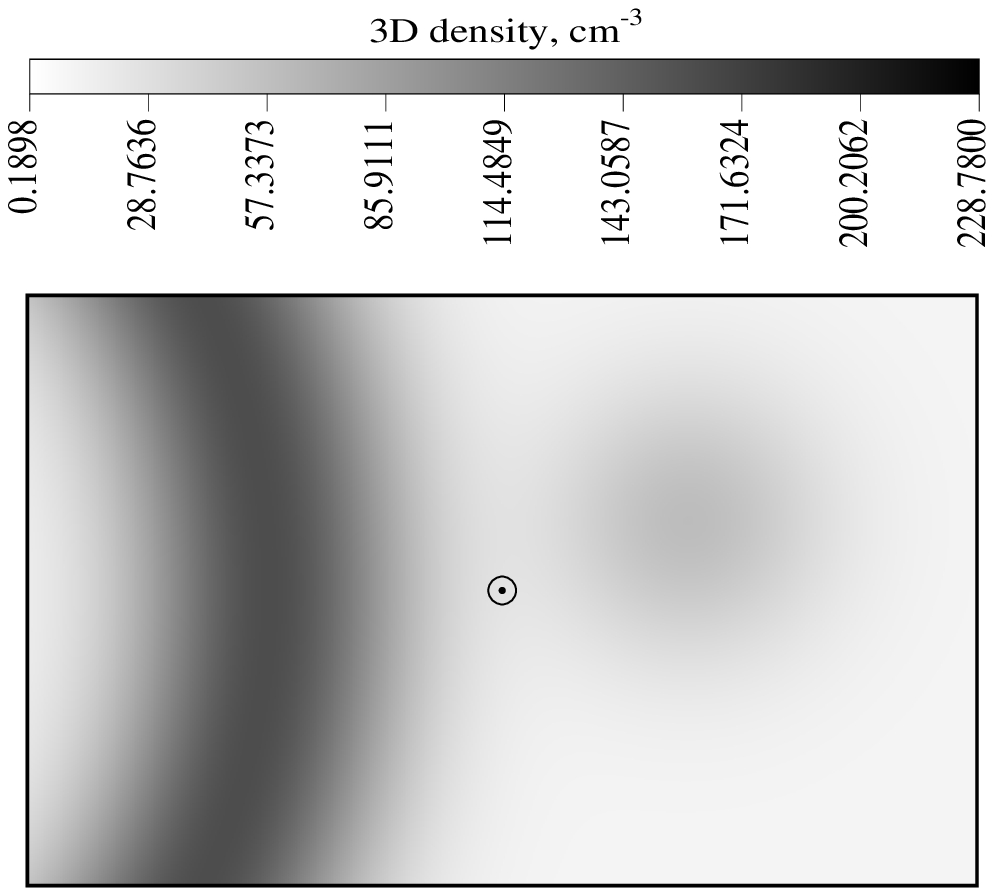}

\figcaption{Simulated density map used to gauge the effect of changing the
form of the entropy used in the reconstruction.  The filled black dots
represent the location of the stars whose intervening columns are used in each
map reconstruction.  The observer location is shown by the sun symbol at the
map center.
\label{f03}}

\clearpage
\plotone{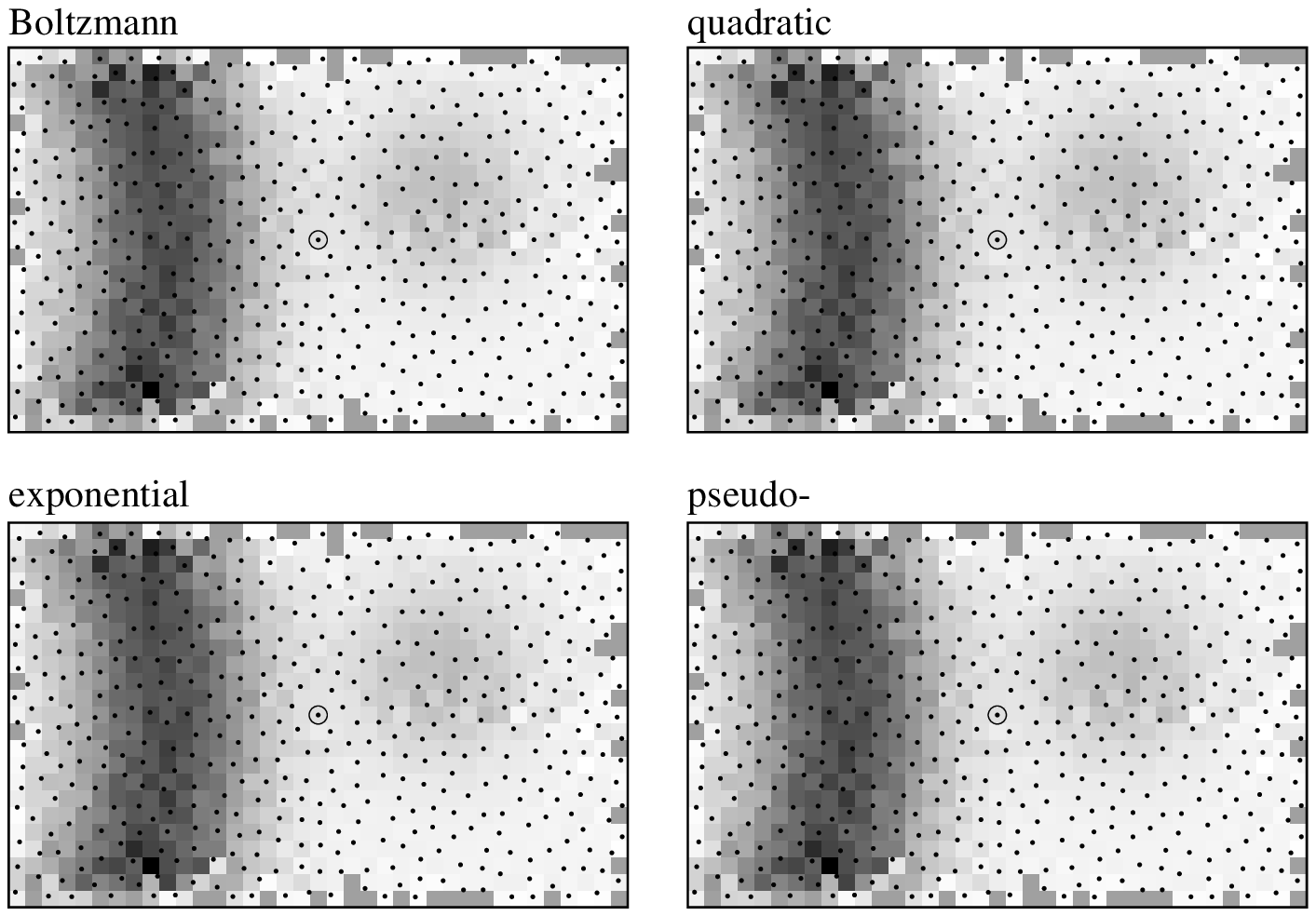}

\figcaption{Maximum entropy reconstructions of the density field shown in
Figure~\ref{f03} using four different entropy functions.  The maps are nearly
indistinguishable.
\label{f04}}

\clearpage
\plotone{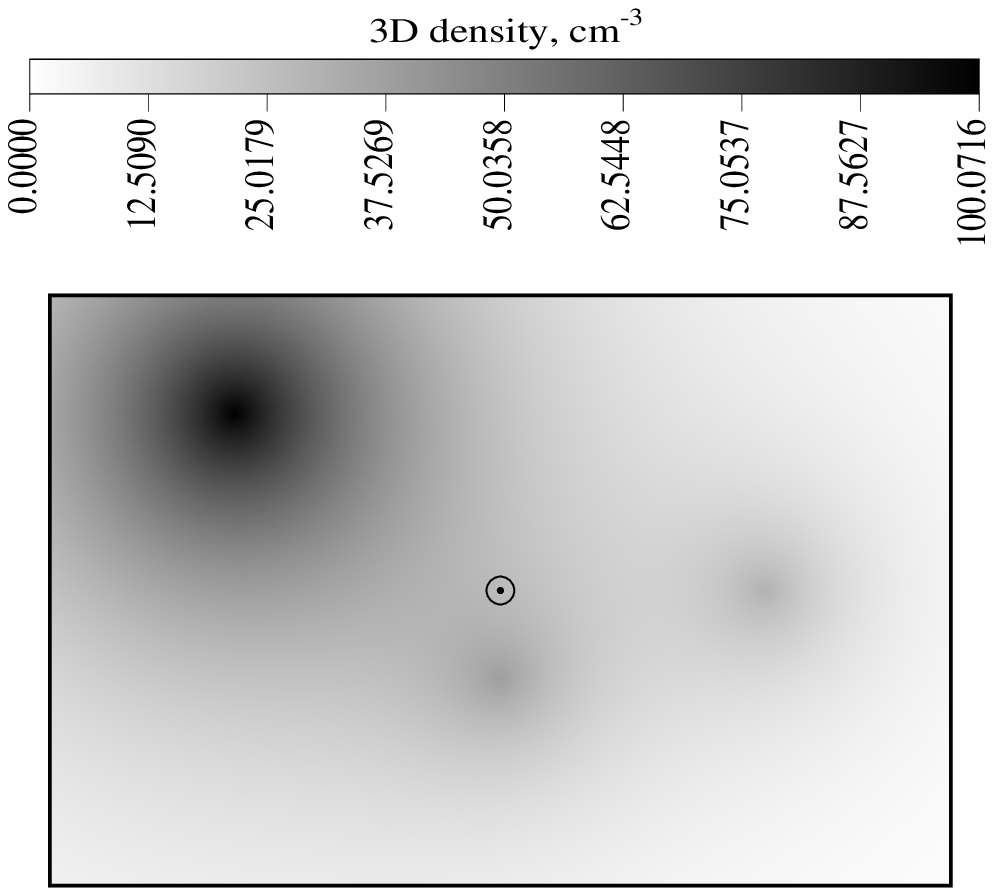}

\figcaption{Simulated density field used to gauge the effect of modulating the
cell size for a given stellar sampling density (Figure~\ref{f06}) and for
modulating the stellar sampling density using an appropriate cell size
(Figure~\ref{f07}).
\label{f05}}

\clearpage
\plotone{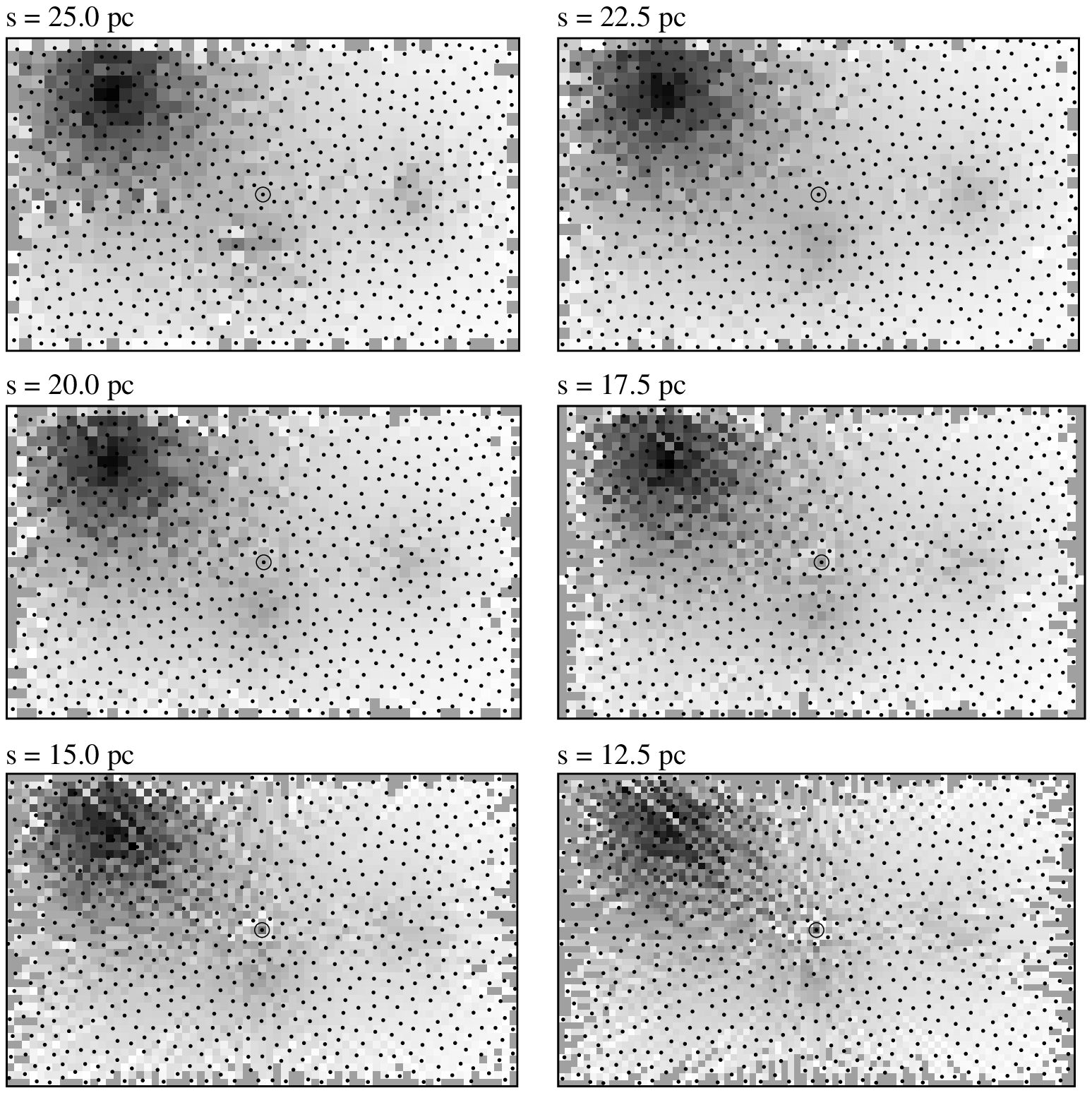}

\figcaption{Maximum entropy reconstructions of the density field shown in
Figure~\ref{f05} using a range of cell sizes.
\label{f06}}

\clearpage
\plotone{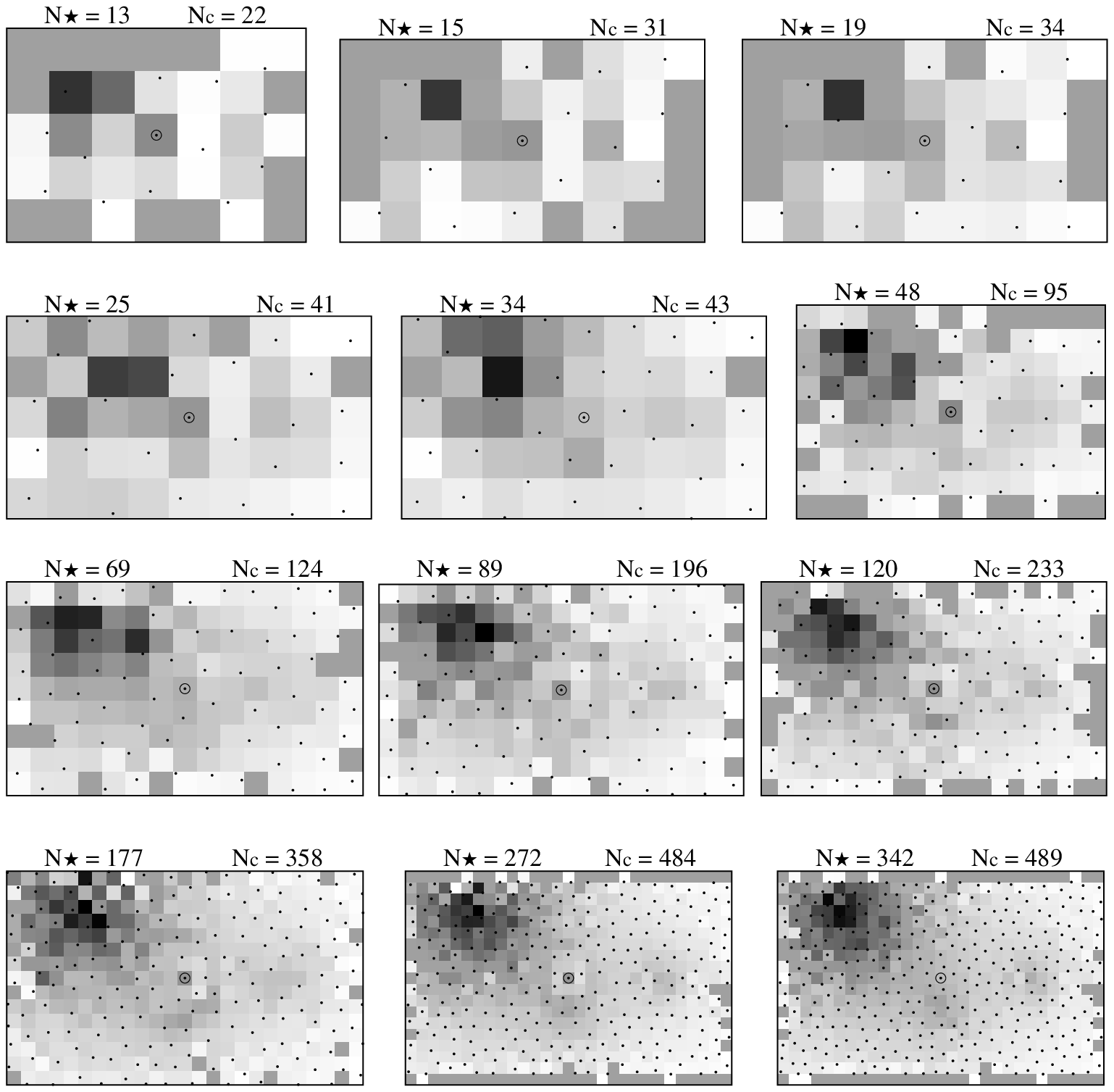}

\figcaption{Maximum entropy reconstructions of the density field shown in
Figure~\ref{f05} using a range of stellar sampling densities and roughly
optimized for cell size.
\label{f07}}


\clearpage
\begin{deluxetable}{llcccc}
\tablewidth{500pt}
\tablecaption{Maximum entropy reconstructions of the simulated data set shown
in Figure~\ref{f03}, for four different entropy functions.  Relative errors
are shown in parentheses. \label{t01}}
\tablehead{
\colhead{entropy}             &
\colhead{functional}          &
\colhead{$\lan N_{mer}-N \ran$}       &
\colhead{$\sqrt{\lan ( N_{mer}-N )^2 \ran}$} &
\colhead{$\lan n_{mer}-n \ran$}       &
\colhead{$\sqrt{\lan ( n_{mer}-n )^2 \ran}$} \\
\colhead{type}     &
\colhead{form}     &
\colhead{(\acmcm)}   &
\colhead{(\acmcm)}   &
\colhead{(\acmcmcm)} &
\colhead{(\acmcmcm)} \\
}
\startdata
Boltzmann      & $S = -\sum \, n_i \log{n_i}$ &
$7.68\tu{16}$  & $6.41\tu{17}$                & $-2.90$     & 16.4 \\
&&(0.000 003)  & (0.000 017)                  & ($-$0.0697) & (0.277) \\
& & & & & \\
quadratic      & $S = -\sum \, n_i^2$         &
$2.56\tu{17}$  & $1.53\tu{18}$                & $-2.90$     & 16.4 \\
&&(0.000 003)  & (0.003 028)                  & ($-0.0697$) & (0.277) \\
& & & & & \\
exponential    & $S = -\sum \, e^{\,n_i}$     &
$7.34\tu{17}$  & $1.74\tu{18}$                & $-2.90$     & 16.4 \\
&&(0.000 025)  & (0.000 049)                  & ($-$0.0697) & (0.277) \\
& & & & & \\
pseudo-        & $S = \sum\,n_i\,e^{\,-n_i}$  &
$2.84\tu{17}$  & $1.96\tu{18} $               & $-2.82$     & 16.2 \\
&&(0.000 003)  & (0.003 031)                  & ($-0.0657$) & (0.265) \\
\enddata
\end{deluxetable}

\clearpage
\begin{deluxetable}{cccccc}
\tablecaption{Maximum (quadratic) entropy reconstructions of the simulated data
set shown in Figure~\ref{f05}, for variable cell size $s$.  Relative errors are
shown in parentheses.  \label{t02}}
\tablehead{
\colhead{s}        &
\colhead{$N_c$}    &
\colhead{$\lan N_{mer}-N \ran$}       &
\colhead{$\sqrt{\lan ( N_{mer}-N )^2 \ran}$} &
\colhead{$\lan n_{mer}-n \ran$}       &
\colhead{$\sqrt{\lan ( n_{mer}-n )^2 \ran}$} \\
\colhead{(pc)}     &
\colhead{    }     &
\colhead{(\acmcm)}   &
\colhead{(\acmcm)}   &
\colhead{(\acmcmcm)} &
\colhead{(\acmcmcm)} \\
}
\startdata
 25.0 & 979         &
$-1.22\tu{19}$  & $3.42\tu{19}$      & $-0.917$    & 6.41 \\
&&(-0.000 716) & (0.001 85)         & ($-0.0535$) & (0.259) \\
\\
 22.5 & 1178        &
$-1.61\tu{17}$  & $9.96\tu{17}$      & $-0.865$    & 5.43 \\
&&(0.000 011)  & (0.000 040)        & ($-0.0505$) & (0.218) \\
\\
 20.0 & 1484        &
$-3.35\tu{17}$  & $7.05\tu{20}$      & $-1.23$     & 6.41 \\
&&(0.000 014)  & (0.000 028)        & ($-0.0672$) & (0.255) \\
\\
 17.5 & 1888        &
$8.09\tu{16}$   & $1.39\tu{18}$      & $-1.39$     & 7.03 \\
&&(0.000 006)  & (0.000 040)        & ($-0.0686$) & (0.264) \\
\\
 15.0 & 2512        &
$1.79\tu{17}$   & $1.07\tu{18}$      & $-1.65$     & 7.75 \\
&&(0.000 005)  & (0.000 037)        & ($-0.0807$) & (0.294) \\
\\
 12.5 & 3477        &
$1.36\tu{19}$   & $3.83\tu{19}$      & $-1.97$     & 8.58 \\
&&(0.000 959)  & (0.003 37)         & ($-0.0919$) & (0.325) \\
\enddata
\end{deluxetable}

\clearpage
\begin{deluxetable}{ccccccc}
\tablecaption{Maximum (quadratic) entropy reconstructions of the simulated data
set shown in Figure~\ref{f05}, for variable $N_{\star}$ and optimized cell
size.  Relative errors are shown in parentheses.  \label{t03}}
\tablehead{
\colhead{$N_{\star}$} &
\colhead{$s$}         &
\colhead{$N_c$}       &
\colhead{$\lan N_{mer}-N \ran$}       &
\colhead{$\sqrt{\lan ( N_{mer}-N )^2 \ran}$} &
\colhead{$\lan n_{mer}-n \ran$}       &
\colhead{$\sqrt{\lan ( n_{mer}-n )^2 \ran}$} \\
\colhead{    }     &
\colhead{(pc)}     &
\colhead{    }     &
\colhead{(\acmcm)}   &
\colhead{(\acmcm)}   &
\colhead{(\acmcmcm)} &
\colhead{(\acmcmcm)} \\
}
\startdata
 13               & 190              & 22           &
$ 7.692\tu{17}$   & $ 2.773\tu{18}$  & $-$2.387     & 10.06       \\
&&&(0.000 031)    & (0.000 113)      & ($-$0.2897)  & (0.7598)    \\
\\
 15               & 140              & 31           &
$ 0.0000\tu{17}$  & $0.0000\tu{18}$  & $-$2.730     & 9.151       \\
&&&(0.000 000)    & (0.000 000)      & ($-$0.1166)  & (0.6934)    \\
\\
 19               & 140              & 34           &
$ 0.0000\tu{17}$  & $0.0000\tu{18}$  & $-$1.387     & 7.258       \\
&&&(0.000 000)    & (0.000 000)      & ($-$0.1039)  & (0.5876)    \\
\\
 25               & 125              & 41           &
$ 0.0000\tu{17}$  & $0.0000\tu{18}$  & $-$2.540     & 12.80       \\
&&&(0.000 000)    & (0.000 000)      & ($-$0.1003)  & (0.4770)    \\
\\
 34               & 120              & 43           &
$ 0.0000\tu{17}$  & $0.0000\tu{18}$  & $-$0.5711    & 8.198       \\
&&&(0.000 000)    & (0.000 000)      & ($-$0.01688) & (0.2676)    \\
\\
 48               & 80               & 95           &
$ 0.0000\tu{17}$  & $0.0000\tu{18}$  & $-$2.826     & 10.88       \\
&&&(0.000 000)    & (0.000 000)      & ($-$0.1266)  & (0.3787)    \\
\\
 69               & 70               & 124          &
$ 0.0000\tu{17}$  & $0.0000\tu{18}$  & $-$2.687     & 10.74       \\
&&&(0.000 000)    & (0.000 000)      & ($-$0.1169)  & (0.3426)    \\
\\
 89               & 55               & 196          &
$ 0.0000\tu{17}$  & $0.0000\tu{18}$  & $-$2.544     & 9.424       \\
&&&(0.000 000)    & (0.000 000)      & ($-$1381)    & (0.3753)    \\
\\
 120              & 50               & 233          &
$ 0.05\tu{17}$    & $0.0258\tu{18}$  & $-$1.980     & 8.327       \\
&&&(0.000 001)    & (0.000 003)      & ($-$0.09216) & (0.3169)    \\
\\
 177              & 40               & 358          &
$ 0.011\tu{17}$   & $0.0106\tu{18}$  & $-$1.692     & 8.556       \\
&&&(0.000 000)    & (0.000 001)      & ($-$0.08350) & (0.2961)    \\
\\
 272              & 35               & 484          &
$ 2.217\tu{17}$   & $1.907\tu{18}$   & $-$1.274     & 7.527       \\
&&&(0.000 007)    & (0.000 029)      & ($-$0.05995) & (0.2609)    \\
\\
 342              & 35               & 489          &
$-1.447\tu{17}$   & $18.60\tu{18}$   & $-$1.178     & 6.801       \\
&&&($-$0.000 041) & (0.000 385)      & ($-$0.0597)  & (0.2549)    \\
\\
\enddata
\end{deluxetable}

\end{document}